\documentclass[
reprint,
superscriptaddress,
 amsmath,amssymb,
 aps,
]{revtex4-2}

\usepackage{graphicx}% Include figure files
\usepackage{dcolumn}% Align table columns on decimal point
\usepackage{bm}% bold math
\newcommand{\TRC}{TianQin Research Center for Gravitational Physics \& School of Physics and Astronomy, Sun Yat-sen University (Zhuhai Campus), Zhuhai 519082, P. R. China}

\newcommand{\CGE}{Center for Gravitational Experiments, School of Physics, MOE Key Laboratory of Fundamental Physical Quantities Measurement \& Hubei Key Laboratory of Gravitation and Quantum Physics, PGMF, Huazhong University of Science and Technology, Wuhan 430074,  P. R. China}

\begin{document}

\preprint{APS/123-QED}

\title{Effect of Earth-Moon's gravity on TianQin's range acceleration noise}

% \thanks{A footnote to the article title}

\author{Xuefeng Zhang}
% \altaffiliation[Also at ]{Physics Department, XYZ University.}
\email{zhangxf38@sysu.edu.cn}
 \affiliation{\TRC}

\author{Chengjian Luo}
\email{luochj5@mail2.sysu.edu.cn}
 \affiliation{\TRC}

\author{Lei Jiao}
 \affiliation{\TRC}

\author{Bobing Ye}
 \affiliation{\TRC}

\author{Huimin Yuan}
 \affiliation{\TRC}

\author{Lin Cai}
\email{cailin@hust.edu.cn}
 \affiliation{\CGE}

\author{Defeng Gu}
 \affiliation{\TRC}

\author{Jianwei Mei}
 \affiliation{\TRC}

\author{Jun Luo}
 \affiliation{\TRC}
 \affiliation{\CGE}

% \collaboration{MUSO Collaboration}%\noaffiliation

\date{\today}% It is always \today, today,

\begin{abstract}
TianQin is a proposed space gravitational-wave detection mission using circular high Earth orbits. The geocentric concept has raised questions about the disturbing effect of the nearby gravity field of the Earth-Moon system on the highly-sensitive intersatellite ranging measurements. Here we examine the issue through high precision numerical orbit simulation with detailed gravity-field models. By evaluating range accelerations between distant free-falling test masses, the study shows that the majority of the Earth-Moon's gravity disturbances are not in TianQin's detection frequency band above $10^{-4}$ Hz, and hence present no showstoppers to the mission. 
\end{abstract}

\maketitle

%\tableofcontents

%%%%%%%%%%%%%%%%%%%%%%%%%%%%%%%%%%%%%%%%%%%%%%%%%%%%%%%%%%%%%%%%%%%%

\section{\label{sec:level1} Introduction}

The current TianQin design assumes high Earth orbits with an orbital radius of $10^5$ km \cite{Luo2016}. The nearly equilateral-triangle constellation stands almost vertically to the ecliptic. High precision laser ranging interferometry tracks distance changes between well-protected test masses (TM) in separate drag-free controlled satellites, within a preliminary frequency range of $10^{-4} - 1$ Hz. Proximity to the Earth has certain benefits such as lower launch cost, shorter transfer duration, easier communication, availability of GNSS (Global Navigation Satellite System), etc. Other geocentric mission concepts \cite{NASA2012} include OMEGA \cite{Hiscock1997,Hellings2011}, GEOGRAWI/gLISA \cite{Tinto2015}, GADFLI \cite{McWilliams2011}, B-DECIGO \cite{Kawamura2018}, etc. TianQin's orbit is different from them in both the orbital radius and orientation.

Space-based gravitational-wave (GW) detectors are subject to various influences from the surrounding environment, including gravity field, thermal radiation, plasma, magnetic field, solar-wind particles, galactic cosmic rays, micrometeorites, etc., that are existential in outer space. Environmental effects can strongly affect performance and lifetime of the sciencecraft. Quite prominently, the space gravity-field environment, encompassing gravitational perturbations from the central and other celestial bodies, plays an important role. It is particularly the case for geocentric missions due to their closeness to the Earth and the Moon. More specifically, the effects on TianQin are two-fold. On large scales, the perturbations distort the nominal equilateral triangle of the constellation. The resulting unequal and time-varying arm-lengths have far-reaching implications on science payload design and data processing strategies (e.g., \cite{Folkner1997,Tinto2014}). The distortion can be reduced by orbit optimization \cite{Ye2019,Tan2020} and control to meet the stability requirements of the science payloads. On small scales, the perturbations impinge on TMs' geodesic motion under nearly pure gravity. Since space-based detectors accurately measure arm-length variations between TMs, they respond not only to GWs (radiation zone), but equally well to Newtonian gravity fields (near zone), in targeted frequency bands. Appearing as environmental noise, the latter should be avoided or mitigated. 

Ideally, gravitational perturbations in space should only manifest as long-term and slow changes in inter-spacecraft displacement measurements. If there exists a proper separation in the frequencies of gravity-field fluctuations and GWs, then the GW signals, superimposed on top of a smooth and slow-varying background, can be extracted (\cite{Barke2015}, Sec 2.1.1). Therefore the GW detection relies heavily on the ``quietness'' of the ambient gravity-field environment in the measurement band. 

The problem of environmental gravity disturbances were recognized early on in designing ground-based detectors \cite{Weiss1972}, and hence is not unique for space missions, where the problem is thought to be much less severe. In ground-based detectors, Newtonian or gravity-gradient noise caused by terrestrial gravity fluctuations poses a limitation to sensitivity improvement below $\sim 10$ Hz \cite{Harms2019}. Multiple strategies have been developed to effectively mitigate such noise, and the techniques have a major influence on designing next-generation ground-based detectors.

If not handled properly in space GW detection, disturbing gravity fields may induce excessive ``orbital noise'' that encroaches on the sensitivity curve, causing a situation somewhat similar to galactic foreground noise in lower frequencies \cite{Nelemans2009}. The potential risk has drawn attentions for TianQin, and may raise concern for other geocentric concepts as well. With regard to LISA \cite{LISA2017}, the majority of the effect is expected to be out of the sensitive frequency band because of its heliocentric yearly orbits and being placed far away from the Earth-Moon system ($\sim 20^\circ$ trailing angle, $\sim 5\times 10^7$ km).

In general, gravity disturbances in space constitute an important potential noise source for inter-spacecraft measurements. In this work, we aim to determine the amplitudes and frequencies of the disturbances for TianQin's orbit, and quantitatively evaluate the impact on TianQin's acceleration noise requirement. The forward modeling takes into account a variety of main gravitational perturbations including the gravity fields of the Earth (static and tidal), the Moon, and the Sun, as well as other solar-system bodies. It requires realistic and accurate orbit propagation that is also used in performance assessment and data analysis of gravity mapping missions such as GRACE \cite{Tapley2004a}, GRACE Follow-On \cite{Abich2019}, GOCE \cite{Floberghagen2011}, and GRAIL \cite{Konopliv2013}. However, for TianQin, a problem with insufficiency of double precision arithmetic has emerged owing to the high measurement accuracy requirement over the long baseline. To tackle the issue, an earlier attempt was made in \cite{Liu2018}, where analytical expansions of perturbed orbits were derived. Unfortunately, the approach cannot handle complicated gravity field models, and only the Earth's static gravity field was considered without realistic Earth's rotation (precession, nutation, etc), the Earth's tides, and third bodies. It motivated us to take a fully numerical approach to be shown in this paper. For other works regarding environmental magnetic and plasmic effects on TianQin, one can refer to, e.g., \cite{Lu2020a,Su2020,Lu2020b}.

This is our third paper of the concept study series on TianQin's orbit and constellation. It is based on the previous work of orbit optimization and constellation stability \cite{Ye2019,Tan2020}, and shifts the attention to small-scale orbital motion through much refined simulation. The paper is organized as follows. In Section \ref{sec:observable}, three types of intersatellite observables are analysed, and the range acceleration is chosen for evaluating the impact. In Section \ref{sec:simulation}, we describe the high precision orbit propagator, detailed force models, and orbital parameters used in the assessment. Section \ref{sec:result} presents the amplitude spectral density (ASD) results of the calculated range accelerations. At the end, the conclusions are made in Section \ref{sec:conclusion}. 

%%%%%%%%%%%%%%%%%%%%%%%%%%%%%%%%%%%%%%%%%%%%%%%%%%%%%%%%%%%%%%%%%%%%

\section{\label{sec:observable} Observables and Criteria}

For the evaluation purpose, the numerical simulation should provide an observable accuracy better than the instrumental measurement noise level. The selectable intersatellite observables include the (instantaneous) range, range-rate, and range acceleration. Mathematically, they are interchangeable by differentiation and integration. But their numerical calculations require different computational resources. Here we estimate the magnitudes of their numerical ranges (numbers of significant digits required) for TianQin. First, the range between two satellites is given by 
\begin{equation}
    \rho = |\mathbf{r}_2 - \mathbf{r}_1|, 
\end{equation}
where $\mathbf{r}_{1,2}$ denotes the position vector of each satellite relative to the Earth's center. Taking the baseline $1.7\times 10^8$ m and the displacement measurement noise $1\times 10^{-12}$ m/Hz$^{1/2}$ \cite{Luo2016}, the numerical representation of the range observable requires at least 20 digits, exceeding the 16 digits of the double-precision format (64 bits). Second, the range rate reads
\begin{equation}
    \dot\rho = \hat{\mathbf{e}}_{12} \cdot (\dot{\mathbf{r}}_2 - \dot{\mathbf{r}}_1),
\end{equation}
with the unit vector $\hat{\mathbf{e}}_{12} = (\mathbf{r}_2 - \mathbf{r}_1)/\rho$. The relative velocities between the TianQin satellites is expected to be within $\pm 5$ m/s \cite{Ye2019}. Taking the range-rate measurement noise $5\times 10^{-14}$ m/s/Hz$^{1/2}$ ($\sim 2\pi f\times 10^{-12}$ m/Hz$^{1/2}$ at the crossover frequency $f\sim 10^{-2}$ Hz of the displacement and residual acceleration noises \cite{Luo2016}), the dynamical range of $\dot \rho$ takes up about 15 digits. Third, differentiating the range rate yields the equation for the range acceleration: 
\begin{equation} \label{eq:rhodd}
    \ddot\rho = \hat{\mathbf{e}}_{12} \cdot (\ddot{\mathbf{r}}_2 - \ddot{\mathbf{r}}_1) + \frac{1}{\rho} \left(|\dot{\mathbf{r}}_2 - \dot{\mathbf{r}}_1|^2 - \dot\rho^2 \right), 
\end{equation}
where, on the right-hand side, the first term represents projected differential acceleration, and the second term centrifugal acceleration. The gravitational acceleration of one TianQin satellite is in the order of $10^{-2}$ m/s$^2$. This is 13 order of magnitude greater than the residual acceleration noise level of one TM, i.e., $1\times 10^{-15}$ m/s${^2}$ \cite{Luo2016}. For either $\dot\rho$ or $\ddot\rho$, if one takes into account that numerical errors compounded over time may occupy 2-3 digits, and redundant numerical accuracy another 1-2 digits, then the requirement would exceed 16 digits. Therefore, the commonly used double-precision arithmetic is insufficient in representing the intersatellite observables, and the associated roundoff error becomes a bottleneck for precision improvement (cf. Fig. \ref{fig:ASD_circle}). 

Among the three observables, the range acceleration appears more favorable for taking up less digits in numerical computation. In the frequency domain, acceleration and displacement can be easily converted. For evaluating gravity disturbances in space, we henceforward adopt the range acceleration as the main observable (cf. \cite{Mueller2017}), and directly compare its ASD with the intersatellite residual acceleration noise requirement $\sqrt{2}\times 10^{-15}$ m/s$^2$/Hz$^{1/2}$ at $10^{-4}-10^{-2}$ Hz as the criteria, which is simply $\sqrt{2}$ of the residual acceleration noise of a single TM (cf. \cite{Armano2016}). Note that this flat noise requirement is preliminary and expected to be relaxed near $10^{-4}$ Hz in the future \cite{Luo2016,LISA2017}. 

%%%%%%%%%%%%%%%%%%%%%%%%%%%%%%%%%%%%%%%%%%%%%%%%%%%%%%%%%%%%%%%%%%%%

\section{\label{sec:simulation} Simulation and Force Models}

The evaluation requires careful calculation and modeling of satellite orbits and gravity fields. The accuracy of numerical integration must surpass the noise requirement $\sqrt{2}\times 10^{-15}$ m/s$^2$/Hz$^{1/2}$ of the range acceleration observable by at least one order of magnitude. The force modeling should be sufficiently detailed and up-to-date to reflect as many significant gravity disturbances as possible, particularly those that may enter the detection band. 

\subsection{Quadruple Precision Orbit Propagation}

\begin{figure}[ht]
\includegraphics[width=0.45\textwidth]{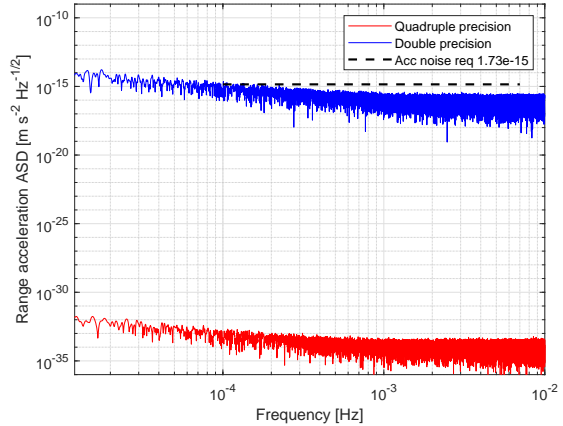}
\caption{\label{fig:ASD_circle} The ASD of the range acceleration $\ddot \rho$ between two satellites in circular orbits of the radius $10^5$ km and separated by $120^\circ$ in phase. The orbits are integrated with a constant step size of 50 s and under the central force of the Earth's point mass. The roundoff error of the quadruple precision arithmetic is at the level of $10^{-33}$ m/s$^2$/Hz$^{1/2}$ (at $10^{-4}$ Hz). The curve tilts up toward low frequencies due to accumulation of roundoff errors over time. For comparison, the roundoff error of double precision is also shown, but at a much higher level of $10^{-15}$ m/s$^2$/Hz$^{1/2}$ (at $10^{-4}$ Hz), hence not sufficient for the accuracy requirement. }
\end{figure}

There exist a few strategies to tackle the inadequacy of double precision. A straightforward way is by extending to 34 significant digits with quadruple precision arithmetic (128 bits). The potential downsides are low execution speed and heavy programming workload. Following this ``brute force'' approach, the TQPOP (TianQin Quadruple Precision Orbit Propagator) program based on MATLAB has been developed so as to evaluate the range acceleration at $<10^{-15}$ m/s$^2$/Hz$^{1/2}$ levels. The quadruple precision data type is applied to all the necessary aspects of the program, including parameter inputs, ephemeris data outputs, reference frame transformations, time conversion, numerical integration, force models, etc. For the nearly circular high orbits, the integrator uses the 8th-order embedded Prince-Dormand (DP87) method \cite{Prince1981} with a constant step size of 50 seconds (Nyquist frequency $10^{-2}$ Hz). The algorithm provides a relative truncation error of $<10^{-20}$ (more than 20 significant digits) in both satellite positions and velocities. Thereby the range acceleration error is estimated to be $<10^{-22}$ m/s$^2$ and well below $10^{-15}$ m/s$^2$. The roundoff error due to finite digits is approximately $10^{-33}$ m/s$^2$/Hz$^{1/2}$, and no longer poses a limiting factor (see Fig. \ref{fig:ASD_circle}), which otherwise would overwhelm gravity field signals in the case of double precision. To mitigate low efficiency of quadruple precision calculations, great effort was made on optimizing code execution to have reduced the run time significantly. Other quadruple precision orbit simulations can be found in, e.g., \cite{McCullough2015,Woeske2016}. 

\subsection{Detailed Force Models}

As the satellites are drag-free controlled, we only consider pure free-fall orbits of the TMs in order to focus on the gravitational perturbative effects. Excluding non-gravitational forces, the force models implemented are summarized in Table \ref{tab:models}. The types of the gravity field models are comparable with those used in the Earth's gravity field determination in satellite missions such as GRACE \cite{Tapley2004a,Tapley2004b} and GOCE \cite{Floberghagen2011}.

The solar system ephemeris uses DE430 \cite{Folkner2014} including all eight planets and the Moon. The effect from the main belt asteroids is estimated not to enter LISA's detection band \cite{Bronicki2018} , nor TianQin's due to the shorter arm-length. Hence they are not included in the simulation.

For the Earth's orientation, the International Astronomical Union (IAU) 2006 precession and IAU 2000A nutation models \cite{Petit2010} are used with the help of the Standards Of Fundamental Astronomy (SOFA) software collection \cite{SOFA}. The Earth's polar motion adopts the IERS Earth Orientation Parameters (EOP) 14 C04 data series \cite{EOP}.

Earth's non-spherical static gravity field is provided by the EGM2008 model \cite{Pavlis2012}, following the recommendation of IERS (2010) \cite{Petit2010}. The normalized spherical harmonic coefficients ($\overline C_{nm}$, $\overline S_{nm}$) are kept up to the 12th degree and order. High-degree terms decay rapidly with increasing radius as $1/r^{n+1}$. Our numerical tests and perturbation analysis shows\ that the effect from the 9th-degree gravity field has already dropped below $10^{-15}$ m/s$^2$/Hz$^{1/2}$. The contribution from the 12th degree sinks deeper to the level of $10^{-18}$ m/s$^2$/Hz$^{1/2}$. Hence we deem it safe to truncate at the degree and order 12. 

Temporal variations of the Earth's gravity field are added as corrections to the spherical harmonic coefficients. To model the Earth's tidal effects, we have followed IERS (2010) \cite{Petit2010} and taken into account solid Earth tides (an-elastic), ocean tides, solid Earth pole tide, and ocean pole tide, as specified in Table \ref{tab:models}. The widely-used ocean tide model FES2004 \cite{Lyard2006} includes long-period ($\Omega_1$, $\Omega_2$, $S_a$, $S_{sa}$, $M_m$, $M_f$, $M_{tm}$, $M_{sqm}$), diurnal ($Q_1$, $O_1$, $P_1$, $K_1$), semi-diurnal ($2N_2$, $N_2$, $M_2$, $S_2$, $K_2$), and quarter-diurnal ($M_4$) waves. The coefficients up to the degree and order 10 are used. Additionally, atmospheric tides are incorporated, though their effect is small compared to the solid Earth and ocean tides. The associated model \cite{Biancale2006} consists of the diurnal and semi-diurnal waves $S_1$ and $S_2$ in the highest frequency constituents. The correction is made up to the degree 8 and order 5. The non-tidal temporal gravity changes have been estimated to be orders of magnitude smaller than the static gravity \cite{Gruber2011}, and will be discussed elsewhere. 

\begin{table}[ht]
\caption{\label{tab:models}
The list of force models implemented in the simulation. }
\begin{ruledtabular}
\begin{tabular}{lc}
Models                 & Specifications \\
\hline
Solar system ephemeris & JPL DE430 \cite{Folkner2014} \\
\hline
Earth's precession \& nutation & IAU 2006/2000A \cite{Petit2010} \\
Earth's polar motion           & EOP 14 C04 \cite{EOP} \\
Earth's static gravity field   & EGM2008 ($n=12$) \cite{Pavlis2012} \\
Solid Earth tides              & IERS (2010) \cite{Petit2010} \\
Ocean tides                    & FES2004 ($n=10$) \cite{Lyard2006} \\
Solid Earth pole tide          & IERS (2010) \cite{Petit2010} \\
Ocean pole tide                & Desai (2003) \cite{Petit2010} \\
Atmospheric tides              & Biancale \& Bode (2003) \cite{Biancale2006} \\ % Model N1
\hline
Moon's libration               & JPL DE430 \cite{Folkner2014} \\
Moon's static gravity field    & GL0660B ($n=7$) \cite{Konopliv2013} \\
% Lunar tide                  & McCarthy \& Petit (2003) \cite{McCarthy2003} \\
Sun's orientation              & IAU \cite{Archinal2011}, Table 1 \\
Sun's $J_2$                    & IAU \cite{Archinal2011}, Table 1 \\
\hline
relativistic effect            & post-Newtonian \cite{McCarthy1996} \\
\end{tabular}
\end{ruledtabular}
\end{table}

Moon's liberation varies about $\pm 8^\circ$, and is provided by DE430 (\cite{Folkner2014}, IIE). For the Moon's static gravity field, we use GL0660B \cite{Konopliv2013} up to the degree and order 7, and the effect of the 7th degree and order alone is below $10^{-20}$ m/s$^2$/Hz$^{1/2}$. The model was obtained from the GRAIL (Gravity Recovery and Interior Laboratory) mission \cite{Konopliv2013} with improved low-degree harmonics. The lunar tide is not included, since the effect is quite small and (semi-)monthly periodic, hence out of the detection band, owning to the Moon's tidal locking with the Earth. The values of the Sun's oblateness $J_2$ and orientation are taken from \cite{Archinal2011} (see also \cite{Folkner2014}, Table 9). Moreover, relativistic effect is added as post-Newtonian correction terms to the equations of motion \cite{McCarthy1996}. The effect is slow varying and expected to be outside the detection band.

\subsection{Orbital Parameters}

The initial orbital parameters are given in Table \ref{tab:orbit}. The integration lasts for one observation window of three months \cite{Luo2016}, that is, from 06 Jun. to 04 Sep. 2004 for 90 days, when the orbital plane is facing the Sun within $\pm 45^\circ$. The year 2004 is chosen without preference but to take advantage of the available EOP observation data, which is more accurate than prediction in 2030s. Our tests have shown that the dominant spectral behavior does not depend on a specific year chosen. 

To make the simulation more realistic, we use the optimized initial orbital elements in Table \ref{tab:orbit} that can meet TianQin's constellation stability requirement (e.g., the breathing angles within $60\pm 0.1^\circ$) for three months \cite{Ye2019,Tan2020}. The optimization removes linear drift in the arm-lengths and breathing angles, and prevents the nearly equilateral-triangle constellation from having severe distortion. The initial eccentricities are set to zeros to keep the orbits almost circular. Note that even if one starts with less optimized initial orbital elements (e.g., the nominal values, $a = 10^5$ km, etc), the dominant spectral behavior of three months (cf. Fig. \ref{fig:ASD_total}) will be unaffected. 

\begin{table}[ht]
\caption{\label{tab:orbit} The initial orbital elements of the TianQin constellation in the J2000-based Earth-centered equatorial coordinate system at the epoch 06 Jun. 2004, 00:00:00 UTC for the evaluation purposes. Here $a$ denotes the semimajor axis, $e$ the eccentricity, $i$ the inclination, $\Omega$ the longitude of ascending node, $\omega$ the argument of periapsis, and $\nu^{\rm ini}$ the truly anomaly. }
\begin{ruledtabular}
\begin{tabular}{ccccccc}
  & $a$ & $e$ & $i$ & $\Omega$ & $\omega$ & $\nu^{\rm ini}$ \\ 
\hline
SC1 & $100000.0$ km & 0 & $74.5^\circ$ & $211.6^\circ$ & $0^\circ$ & $30^\circ$ \\
SC2 & $100009.5$ km & 0 & $74.5^\circ$ & $211.6^\circ$ & $0^\circ$ & $150^\circ$ \\
SC3 & $ 99995.0$ km & 0 & $74.5^\circ$ & $211.6^\circ$ & $0^\circ$ & $270^\circ$ \\
\end{tabular}
\end{ruledtabular}
\end{table}

%%%%%%%%%%%%%%%%%%%%%%%%%%%%%%%%%%%%%%%%%%%%%%%%%%%%%%%%%%%%%%%%%%%%

\section{\label{sec:result} Spectral Results}

It should be emphasized that the purpose of this work is to determine the frequency-domain effects (especially $>10^{-4}$ Hz) of various gravity disturbances on the range acceleration observable, and it concerns less about the absolute accuracy of an integrated orbit, which may drift away from true values over long time scales outside the frequency band of interest. 

\subsection{Total Effect}

\begin{figure}[ht]
\includegraphics[width=0.45\textwidth]{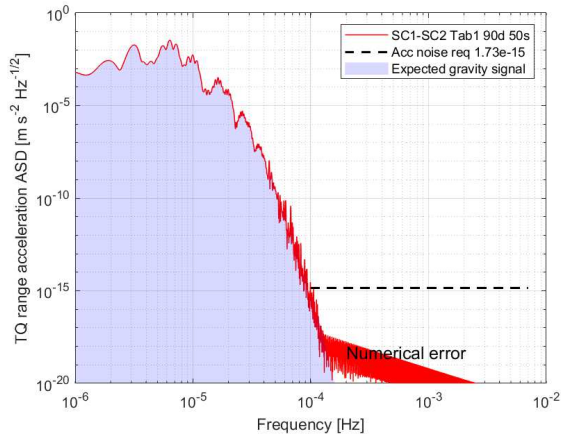}
\caption{\label{fig:ASD_total} The range acceleration ASD of two TianQin satellites SC1 and SC2, calculated in quadruple precision and with the models of Table \ref{tab:models} and step size 50s using 90 days of data. The orbital period 3.6 days corresponds to $3.2\times 10^{-6}$ Hz. The plots for SC1-SC3 and SC2-SC3 are nearly identical to the one shown above. }
\end{figure}

The overall result of the range acceleration ASD is presented in Fig. \ref{fig:ASD_total} for the arm SC1-SC2 using the models of Table \ref{tab:models} assembled together in the simulation. The plots for the other two arms severely overlap with the first one, hence not presented for clarity. 

In the frequency domain, the gravity field signals are dominating below $10^{-4}$ Hz, and roll off rapidly in amplitudes toward high frequencies, and intersect with the lower end of the range acceleration noise requirement at $1\times 10^{-4}$ Hz The steep fall-off roughly follows a power law of $\sim 1.7\times 10^{-15}$ m/s$^2$/Hz$^{1/2}$ $\times (0.1\ \textrm{mHz}/f)^{24}$ near $f = 1\times 10^{-4}$ Hz. The plot demonstrates that the effect of the gravity field models in Table \ref{tab:models} does not enter the detection band $>10^{-4}$ Hz. Note that the slanted part of the ASD curve ($<10^{-17}$ m/s$^2$/Hz$^{1/2}$ and $>10^{-4}$ Hz, marked by ``Numerical error'' in Fig. \ref{fig:ASD_total}) is an artifact of numerical interpolation of the EOP data. 

The frequency-domain behavior somewhat resembles the one in the intersatellite laser ranging measurement result of the GRACE Follow-On mission \cite{Abich2019}, which also shows a steep fall-off, but at a higher frequency ($\sim 4\times 10^{-2}$ Hz) because of its low orbit altitude of approximately 500 km. 

%%%%%%%%%%%%%%%%%%%%%%%%%%%%%%%%%%%%%%%%%%%%%%%%%%%%%%%%%%%%%%%%%%%%

\subsection{Effect Breakdown}

\begin{figure}[ht]
\includegraphics[width=0.45\textwidth]{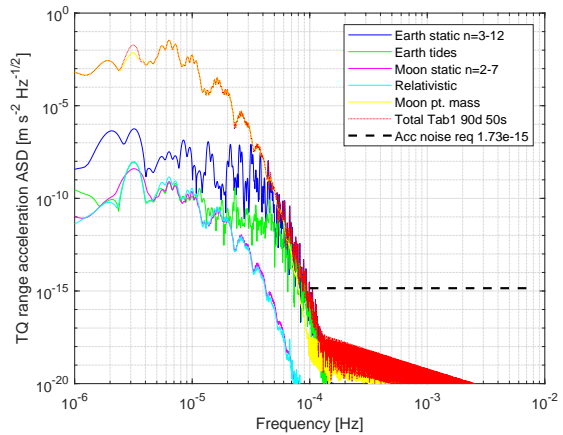}
\caption{\label{fig:ASD_breakdown} Components of the range acceleration ASD of two TianQin satellites SC1 and SC2. The total ASD in Fig. \ref{fig:ASD_total} is duplicated in dotted red curve for comparison. }
\end{figure}

Now we examine various contributions to the total range acceleration ASD of the arm SC1-SC2. The result is presented in Fig. \ref{fig:ASD_breakdown}.

The Earth's non-spherical static gravity field with degrees $n\geq 3$ dominates above $5\times 10^{-5}$ Hz in the total ASD, indicated by the overlapping of the blue and red curves. The effect decreases rapidly toward high frequencies and impinges on the noise requirement at $10^{-4}$ Hz. At the high altitude of TianQin, high-degree harmonics are effectively attenuated.

The contribution from the Moon's non-spherical static gravity field ($n\geq 2$) is minute and only sticks out in the low-frequency region. One may have expected so since the Moon is a slowly rotating body. The same argument also justifies the omission of the lunar tides in the simulation. Nevertheless, the Moon's point mass and its orbital motion play an important role, largely accounting for the total ASD below $4\times 10^{-5}$ Hz.

The effects of relativity and Earth's tidal gravity field (solid Earth, oceanic, pole, and atmospheric, cf. Table \ref{tab:models}) are considerably smaller than the total effect, and both peak at low frequencies away from the detection band. These low frequency disturbances show no significant coupling into high frequencies, and do not induce pronounced range acceleration response above $10^{-4}$ Hz. 

%%%%%%%%%%%%%%%%%%%%%%%%%%%%%%%%%%%%%%%%%%%%%%%%%%%%%%%%%%%%%%%%%%%%

\subsection{Model Errors}

The models inevitably contain errors. To estimate their effect, a straightforward way is to determine whether discrepancies between different models can significantly alter the spectral result in Fig. \ref{fig:ASD_total}. 

\begin{figure}[ht]
\includegraphics[width=0.45\textwidth]{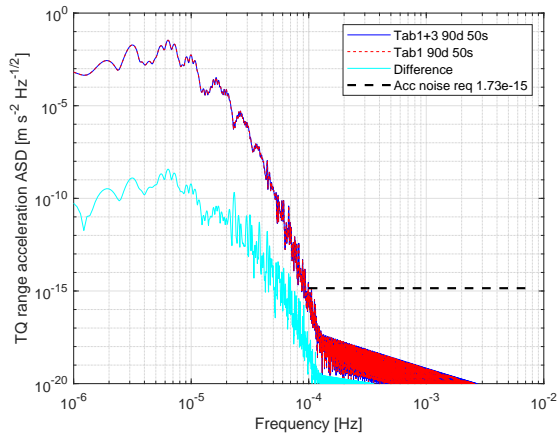}
\caption{\label{fig:ASD_error} The range acceleration ASD of two TianQin satellites SC1 and SC2 using the models of Table \ref{tab:models2} for replacement to Table \ref{tab:models} (Tab1+3, blue). The ASD in Fig. \ref{fig:ASD_total} is duplicated in dashed red curve (Tab1) for comparison. The ASD of their difference $\Delta \ddot\rho$ is marked by cyan curve. }
\end{figure}

For a cross-check, we test another set of gravity field models shown in Table \ref{tab:models2} where several replacements are made to Table \ref{tab:models}. The substitute models are deemed less accurate than the corresponding more recent ones in Table \ref{tab:models}, and thus can mimic model errors (see, e.g., \cite{Loomis2012} ). In Fig. \ref{fig:ASD_error}, both ASD results show good agreement with each other, and the difference $\Delta \ddot\rho$ is well below the noise requirement. In addition, the spectral behavior above $10^{-15}$ m/s$^2$/Hz$^{1/2}$ is also confirmed by running the flight-qualified, open source program GMAT \cite{GMAT} in double precision. Hence the overall frequency-domain behavior appears to be robust, which instills more confidence in the results. 

\begin{table}[ht]
\caption{\label{tab:models2}
The list of replacement force models to Table \ref{tab:models} used for spectrum comparison. }
\begin{ruledtabular}
\begin{tabular}{lc}
Models                 & Specifications \\
\hline
Solar system ephemeris & JPL DE405 \cite{Standish1998} \\
\hline
Earth's precession \& nutation & IAU 1976/1980 \cite{McCarthy1996} \\
Earth's static gravity field   & EGM96 ($n=12$) \cite{Lemoine1998} \\
% Ocean tides                    & EOT08a ($n=10$) \\
\hline
Moon's libration               & JPL DE405 \cite{Standish1998} \\
Moon's static gravity field    & LP165P ($n=7$) \cite{Konopliv2001} \\
\end{tabular}
\end{ruledtabular}
\end{table}

%%%%%%%%%%%%%%%%%%%%%%%%%%%%%%%%%%%%%%%%%%%%%%%%%%%%%%%%%%%%%%%%%%%%

\section{\label{sec:conclusion} Conclusion} 

The TianQin mission, to be deployed in a high Earth orbit, shares technological similarities with low-Earth gravimetry missions using satellite-to-satellite tracking. They diverge on a key point that the Earth's gravity field signals targeted in gravimetry missions become environmental noise in TianQin's GW detection. Hence TianQin must keep a safe distance from the Earth by flying high enough, so as to push the Earth's gravity field interference out of the detection band. This work has been devoted to evaluating and examining this type of effect, and two main conclusions can be drawn here.

1. With the orbital radius of $10^5$ km for TianQin, the current models show that the effect of the Earth-Moon's gravity field dominates at low frequencies, and that the amplitude rolls off rapidly toward high frequencies and intersects with the range acceleration noise requirement ($\sqrt{2}\times 10^{-15}$ m/s$^2$/Hz$^{1/2}$) at $10^{-4}$ Hz, right on the lower end of the preliminary detection band. To provide more context, the gravity field signals from GRACE-FO laser ranging interferometry along a $\sim 200$ km baseline falls off at about $4\times 10^{-2}$ Hz \cite{Abich2019} with an orbit altitude of $\sim 500$ km. 

2. The high-precision numerical simulations help to rule out the majority of the perturbing gravity sources for TianQin, including the Sun's point mass and its $J_2$, the solar system planets' point masses (under their orbital motion), the Earth's static gravity (with its rotation), the Earth's tidal gravity changes (solid Earth, oceanic, pole, and atmospheric), Moon's static gravity, relativistic effect, etc (cf. Table \ref{tab:models}). These effects are slowly varying, not entering the detection band, and present no show-stoppers for TianQin. The Newtonian gravity-field environment at a distance of $10^5$ km from the Earth is expected to be fairly ``quiet'' for TianQin.

The results can provide useful inputs and guidelines to several aspects of the mission concept studies, such as orbit selection, noise reduction, and data processing. For future works, further refined gravity models will be explored to identify other possible noise sources. On another note, the strong low-frequency gravity field signals ($<10^{-4}$ Hz) illustrated in Fig \ref{fig:ASD_total}, which carry long-wavelength gravity information, may find potential applications in geodesy and geophysics \cite{Jiao2020}. This may help to enrich TianQin's secondary science output.

%%%%%%%%%%%%%%%%%%%%%%%%%%%%%%%%%%%%%%%%%%%%%%%%%%%%%%%%%%%%%%%%%%%%

\begin{acknowledgments}
The authors thank Qiong Li, Xiaoli Su, Fan Yang, Hao Zhou, Vitali M\"uller, Gerhard Heinzel, Bo Xu, Shuai Liu, Yi-Ming Hu, Hui-Zong Duan, Lin Zhu, Jinxiu Zhang, Cheng-Gang Shao, Shan-Qing Yang, Hsien-Chi Yeh, and anonymous referees for helpful discussions and comments. The work is supported by the National Key R\&D Program of China (No. 2020YFC2201202). X.Z. is supported by NSFC Grant No. 11805287. 
\end{acknowledgments}

\bibliography{apssamp}% Produces the bibliography via BibTeX.

\end{document}